\documentclass[twocolumn,prb,showpacs]{revtex4}

\usepackage{graphicx}
\usepackage{dcolumn}
\usepackage{bm}

\begin{document}

\title{Pressure effects on the electronic properties
and superconductivity
in the $\beta$-pyrochlore oxides: $A$Os$_2$O$_6$ \\
($A$ = Na, K, Rb, Cs)}

\author{R. Saniz and A. J. Freeman}
\affiliation{Department of Physics and Astronomy,
Northwestern University, Evanston, Illinois 60208-3112, USA}

\date{\today}
                                                                
                                             
\begin{abstract}
We present a first-principles study of the
electronic structure and superconducting parameters of the
compounds $A$Os$_2$O$_6$ ($A$ = Na, K, Rb, and Cs) and at
ambient and applied hydrostatic pressure. We find that the
sensitivity of the
density of states at the Fermi energy, $E_{\rm F}$, and related
electronic properties to
the size of the alkali metal atom as well as to applied
pressure is driven by a van
Hove singularity with energy very close to $E_{\rm F}$.
Further, a computation of the superconducting parameters of these
materials allows us to show that the observed change of $T_c$,
both upon substitution of the alkali metal and under applied 
hydrostatic pressure, can be well understood
within a phonon-mediated pairing scenario. In this regard, we find
that the
correction to the effective electron mass due to spin fluctuations
plays a significant role.

\end{abstract}

\maketitle

\section{Introduction}

During the last decade or so, the search and interest in the
superconductivity of non-Cu based oxides has extended from
an effort to understand the pairing mechanism in the cuprates to
a broader search of superconductivity in materials in which
electron correlations are thought to play a determining role. To
this end, researchers try to exploit a diverse range of factors,
from the orbital degrees of freedom to the crystal
structure of the material. A particularly interesting example
is the recently discovered family of superconducting
Os-oxide $\beta$-pyrochlores {\it A}Os$_2$O$_6$, with $A$ = K, Rb,
and Cs,\cite{yonezawa04a,yonezawa04b,yonezawa04c} which have
superconducting transition temperatures $T_c=9.6$ K, 6.3 K, and
3.3 K, respectively. The pyrochlore structure is a network of
corner sharing tetrahedra and is a geometrically frustrated spin
system if the ions bear a localized magnetic moment interacting
antiferromagnetically with their nearest neighbors.
Further, the Os ions---located at the tetrahedra
vertices---possess a formal oxidation state of $5.5+$
($5d^{2.5}$). As pointed
out by Hiroi and co-workers,\cite{hiroi04} compared to other
transition metal pyrochlore oxides, this places these
materials between Cd$_2$Re$_2$O$_7$ (Re$^{5+}: 5d^2$), which is a
good conductor at low temperatures\cite{hanawa01,jin01}
(becoming superconductor at $\sim 1$ K)
and Cd$_2$Os$_2$O$_7$ (Os$^{5+}: 5d^3$), which is
an insulator at low temperatures and exhibits antiferromagnetic
ordering.\cite{mandrus01} 

While 
the several experimental results\cite{yonezawa04a,yonezawa04b,
yonezawa04c,bruehwiler04,
magishi04,arai04,koda04,khasanov04,muramatsu04}
reported during the past year
provide key information,
the pairing mechanism in these materials is still under debate.
These experiments indicate similarities but also
differences among the compounds with different
alkali metal, tending to single out KOs$_2$O$_6$.
Let us mention briefly some of the observations that should
be taken into account by any proposed pairing mechanism.
Firstly, the very
change in $T_c$ upon substitution of the alkali atom ($A$) may be
initially counterintuitive. Indeed, the negative chemical pressure
leads to an increase of the lattice constant
with the ionic radius of $A$, so that one
may expect $T_c$ to increase because band narrowing should lead
to an increase of the density of states (DOS)
at the Fermi level ($E_{\rm F}$).
As shown by the reported
$T_c$'s above, however, the opposite occurs. In line with these
findings, the change in $T_c$ under applied hydrostatic pressure
is found to be positive initially in all these
materials.\cite{khasanov04,muramatsu04}
This has been
interpreted as suggesting that the pairing mechanism in these
materials is unconventional, or
non-BCS-like.\cite{yonezawa04c} Of interest is the fact that in
KOs$_2$O$_6$ the increase of $T_c$ with pressure reaches a
maximum at 0.56 GPa and tends to vanish gradually at higher
pressures.\cite{muramatsu04,comment0} 

\begin{table*}
\caption{\label{tab1} Structural parameters of the Os-oxide
$\beta$-pyrochlores within the GGA approximation
($T=0$ K). }
\begin{ruledtabular}
\begin{tabular}{ccccccc}
Compound  & $a$ (\AA) & $x$ & $l_{\rm Os-O}$ (\AA) & $\angle$O-Os-O &
$\angle$Os-O-Os & $B$ (GPa) \\
\hline
NaOs$_2$O$_6$ & 10.274 & 0.317 & 1.942 & 88.26 & 138.57 & 113.9 \\
KOs$_2$O$_6$ & 10.298 & 0.316 & 1.944 & 88.54 & 138.98 & 116.6 \\
RbOs$_2$O$_6$ & 10.318 & 0.315 & 1.945 & 88.82 & 139.39 & 118.8 \\
CsOs$_2$O$_6$ & 10.356 & 0.314 & 1.948 & 89.27 & 140.03 & 123.8 \\
\end{tabular}
\end{ruledtabular}
\end{table*}

Further observations are that the temperature
dependence of the resistivity shows an unusual concave
behavior at low temperature
in the case of KOs$_2$O$_6$,\cite{yonezawa04a} while
a $T^2$
behavior is observed\cite{yonezawa04b} just above $T_c$ in the
case of RbOs$_2$O$_6$
and on a larger temperature
interval\cite{yonezawa04c} in the case of CsOs$_2$O$_6$.
Also, nuclear magnetic resonance (NMR) experiments reveal a weak
temperature dependence of the Knight shift of both the $^{39}$K
and $^{87}$Rb nuclei in the corresponding
pyrochlores.\cite{magishi04,arai04} In
the normal state, however, the relaxation rate
divided by temperature ($1/TT_1$) follows the Korringa relaxation
in the case of Rb,\cite{magishi04} or deviates weakly from
it,\cite{arai04} but deviates more
strongly from this behavior in the case of K.\cite{arai04}
In both cases, however, there appears to be evidence for
antiferromagnetic spin fluctuations. Finally,
regarding the superconducting gap,
Magishi and co-workers find that the relaxation rate
in the superconducting state suggests an anisotropic but
nodeless gap\cite{magishi04} in RbOs$_2$O$_6$; at the same
time, Koda and collaborators
interpret their muon spin rotation ($\mu$SR)
study of the magnetic penetration depth
in KOs$_2$O$_6$ as pointing to an anisotropic gap with
nodes.\cite{koda04} 
 
We report here on a first-principles study of the
electronic structure and superconducting parameters of the
compounds $A$Os$_2$O$_6$ ($A$ = Na, K, Rb, and Cs) and
on the effects
of hydrostatic pressure. We find that the main traits
of the electronic structure reported previously\cite{saniz04}
in the case of
KOs$_2$O$_6$ are common to all these materials, with relatively
small qualitative and quantitative changes. The differences
stem essentially from the energy level with respect to
$E_{\rm F}$ of the van Hove singularity
(vHS) with momentum
${\bf k}$ near the center of the $\Gamma$-$L$ line.
In particular, the density of states at the Fermi energy,
$N(E_{\rm F})$, tends to
increase with the size of $A$ because the
vHS is pushed closer to $E_{\rm F}$.
The effect of applied hydrostatic pressure is to push the vHS
away from $E_{\rm F}$. This is very clearly reflected by
the increase or suppression of a
constriction between the two
$\Gamma$-point centered Fermi surface shells,
basically due to a bending of the outer shell that depends
on the proximity of the vHS to $E_{\rm F}$.

We further estimate
$T_c$ with the well-known McMillan-Allen-Dynes
expression,\cite{bennemann72} with the
electron-phonon coupling constant calculated within the crude
rigid muffin-tin approximation (RMTA).\cite{gaspari72,pettifor77}
We also calculate the Stoner
susceptibility enhancement parameter
and estimate the electron-spin coupling constant within the
Doniach-Engelsberg approximation.\cite{doniach66} This allows
us to show that spin fluctuations contribute importantly to
the effective electron mass, significantly reducing $T_c$.
Despite the approximations implicit in these
calculations, we find, remarkably, that the calculated $T_c$
follows rather well the trends observed in experiment,
both upon substitution of the alkali metal and
under hydrostatic pressure. Our results, thus, bring further
support to the electron-phonon coupling description of these
superconductors.\cite{bruehwiler04,magishi04,khasanov04,saniz04}

Section II is devoted to the methodology of our calculations
as well as to structural properties;
in section III, we present and discuss our results in relation
to the experimental findings mentioned above.

\section{Methodology and structural aspects}

We use the highly precise full-potential
linearized augmented plane-wave
(FLAPW)\cite{wimmer81}
implementation of the
density functional approach to the electronic structure and
properties of crystalline solids.
We make our self-consistent calculations within the Perdew,
Burke, and Ernzerhof generalized gradient approximation
(GGA)\cite{perdew97} to the exchange-correlation potential
and include the spin-orbit coupling (SOC) term in the Hamiltonian.
Angular momenta up to $l=8$ are used for both the
charge density in the muffin-tins and the wave functions. The
irreducible part of the Brillouin zone is sampled with a uniform
mesh of 120 {\bf k}-points. The Os $5p$ and K $3p$ states are
treated as valence electrons.

The $\beta$-pyrochlores crystallize in a cubic structure with
space group $Fd\bar3m$. There are 18 atoms in the unit cell:
two $A$ atoms ($8b$), four Os atoms ($16c$), and twelve O
atoms ($48f$). The Os atoms are octahedrally coordinated by six
O atoms. An internal parameter, $x$, fixes the positions of the
latter and thereby also determines the degree of rhombohedral
distortion of the octahedra enclosing the Os atoms.
In all the cases considered, we determine
the lattice constant, $a$,
as well as $x$,
by minimizing the total energy and ensuring the total force on
the O atoms is less than $10^{-4}$ a.u.
The muffin-tin radii used are 2.2 a.u. for the Os ions
and 1.3 a.u. for the O ions.
The corresponding values for the Na, K, Rb, and Cs ions are,
respectively, 2.6, 2.8, 3.0, and 3.1 a.u.

\begin{figure*}
\begin{center}
\includegraphics[width=.9\hsize]{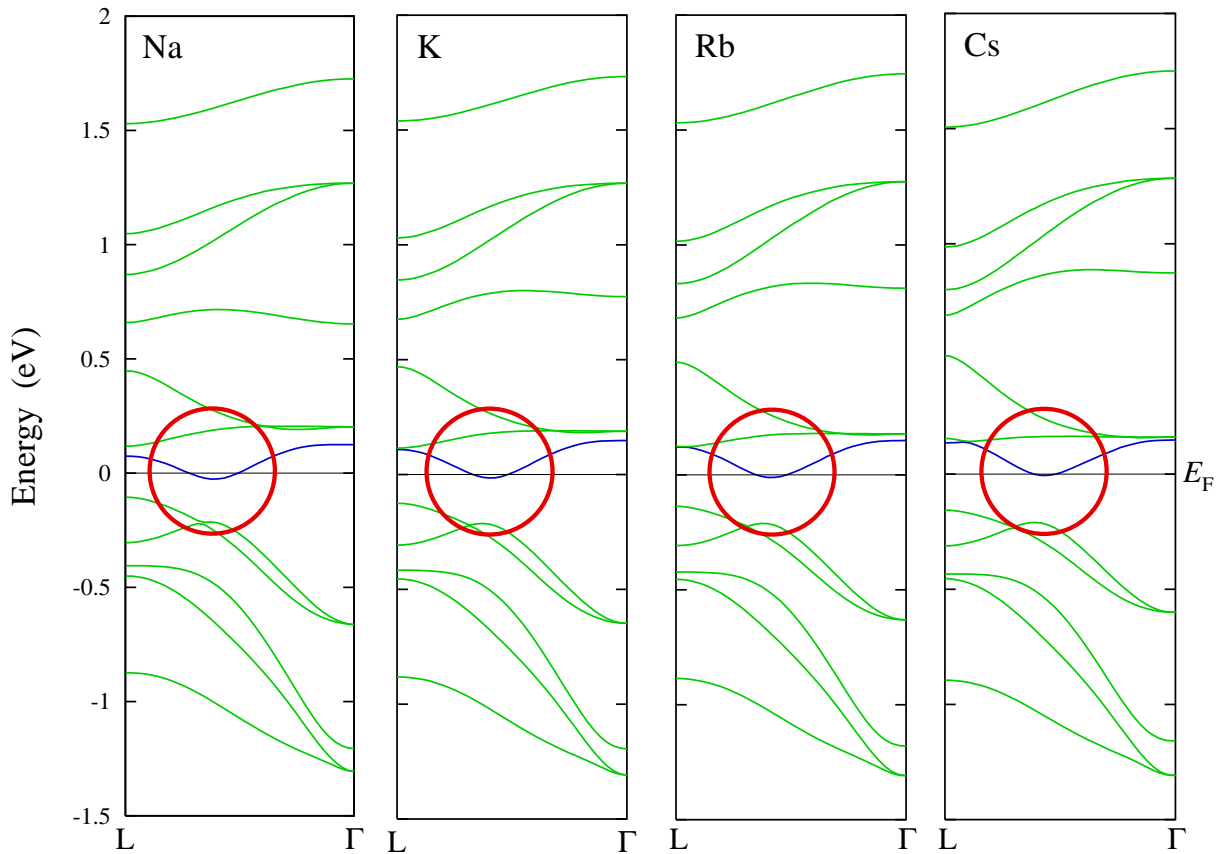}
\end{center}
\caption{\label{fig1} (Color online)
The manifold of 12 energy bands near $E_{\rm F}$
of the superconducting
$\beta$-pyrochlore Os oxides plotted for {\bf k}-points along the
$L$-$\Gamma$ line.
The band crossing $E_{\rm F}$ is highlighted in blue, and the red
circle indicates the location of the van Hove singularity.
It is clearly seen
that the larger the alkali metal, the closer the singularity is
to $E_{\rm F}$.}
\end{figure*}

The values of the calculated structural parameters are reported
in Table~\ref{tab1}, including the smaller Os-O-Os angle
defining the
main rectangular
cross section of the octahedra and the O-Os-O angle characterizing
the staggered Os-O chains on the underlying the pyrochlore lattice.
The calculated ($T=0$ K) lattice constants differ from the
(room temperature) experimental results of Hiroi and co-workers
(see Ref.~\onlinecite{muramatsu04}) by $+1.95$\% for KOs$_2$O$_6$,
$+1.98$\% for RbOs$_2$O$_6$, and $+2.00$\% for CsOs$_2$O$_6$.
Although the compound with Na has not been synthesized---probably
because of its small size---we
have included it in our study to better identify the
trends followed by the different properties upon $A$ substitution.
With respect to the other parameters in
Table~\ref{tab1}, unfortunately the only experimental values
reported are those
of Br\"uhwiler {\it et al.}\cite{bruehwiler04} for RbOs$_2$O$_6$.
In this case, the difference
of the calculated Os-O bond length from
experiment is $+1.8\%$, while the difference for
the internal parameter, and the O-Os-O and Os-O-Os angles are,
notably, all below 0.1\%. From Table~\ref{tab1}, it is clear that
the angle in the Os-O chains increases with decreasing $T_c$.
Assuming a phonon mediating pairing,
it was suggested\cite{bruehwiler04} that this angle plays a role
in determining $T_c$. Future studies of the phonon spectra of
these materials should allow one to verify this interesting point.
In Table~\ref{tab1}, we also give the calculated bulk moduli,
which are necessary to calculate volume changes under
pressure.

\section{Electronic structure and properties}

\subsection{Band structure and density of states}

\begin{figure}
\begin{center}
\includegraphics[width=\hsize]{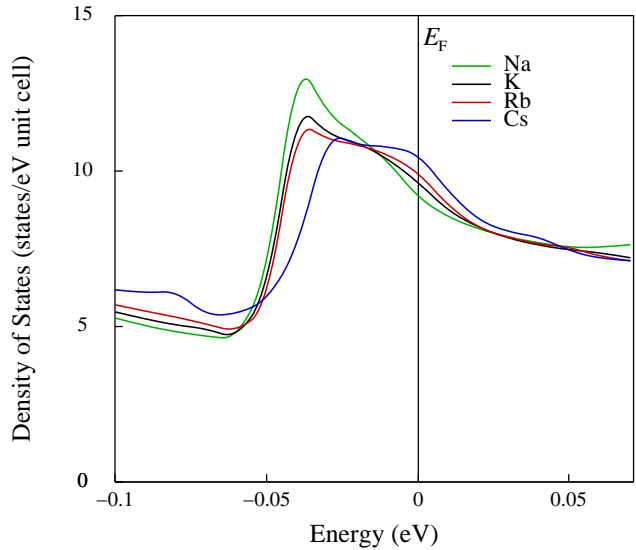}
\end{center}
\caption{\label{fig2} (Color online)
Density of states at $E_{\rm F}$ for the
$\beta$-pyrochlore Os oxides. Although the height of the
van Hove peak itself
tends to decrease with the size the alkali ion, the value of
$N(E_{\rm F})$ increases because the
singularity shifts up in energy toward $E_{\rm F}$.}
\end{figure}

As we reported previously in the case of
KOs$_2$O$_6$,\cite{saniz04}
we find that in all the $\beta$-pyrochlores
the band structure around $E_{\rm F}$ is given by a manifold of
twelve bands arising mainly from Os $5d$ states and
O $2p$ states. The dispersion of the bands is generally the same
for the different compounds, but there is an important difference
near $E_{\rm F}$, which is that the vHS
near the center of the $\Gamma$-$L$ line moves up closer to the
Fermi level as the size of $A$ increases. This is
clearly illustrated in Fig.~\ref{fig1}, where we compare the
energy bands of the different compounds for {\bf k}-points
along the $\Gamma$-$L$. The consequence of
this for DOS is also intriguing.
Indeed, although the peak due to the vHS tends to decrease
for the cases with a larger $A$ ion, 
$N(E_{\rm F})$ increases because the vHS is closer to $E_{\rm F}$.
This is clear from Fig.~\ref{fig2}, where we show a close-up
of the total DOS around $E_{\rm F}$ for the different compounds.
In Table ~\ref{tab2},
where we list the total DOS at $E_{\rm F}$, as well as
the muffin-tin sphere projected DOS for O and Os. For reference,
we also indicate the values of the bare band Sommerfeld
coefficient and of the band Pauli paramagnetic susceptibility.
Comparing with measurements on powder samples,
the specific heat
mass-enhancements, $\gamma_{\rm exp}/\gamma_{\rm b}$, appear to
be 3.3 for KOs$_2$O$_6$
($\gamma_{\rm exp}=19$ mJ/K$^2$ mol Os)\cite{hiroi04}
and 3.7 for RbOs$_2$O$_6$
($\gamma_{\rm exp}=44$ mJ/K$^2$ mol Os).\cite{bruehwiler05}
If the results reported by Muramatsu {\it et al.} are used for
RbOs$_2$O$_6$ and CsOs$_2$O$_6$ (both with
$\gamma\simeq 20$ mJ/K$^2$ mol Os)\cite{muramatsu04}
one finds mass enhancements of 3.4 and 3.2, respectively.
We note, however, that very recently Hiroi and co-workers
reported\cite{hiroi05} 
specific heat measurements on a single crystal sample of
KOs$_2$O$_6$, which, making an estimate similar to the powder sample
case, yields a surprising $\gamma_{\rm exp}$=64.8 mJ/K$^2$ mol Os.
This results in an unusually large
$\gamma_{\rm exp}/\gamma_{\rm b}\simeq 11.4$.\cite{comment1}

\begin{table}
\caption{\label{tab2} Total DOS
(states/eV unit cell) and muffin-tin sphere
projected DOS (states/eV atom) at $E_{\rm F}$.
Also given are the bare band Sommerfeld coefficient
(mJ/K$^2$ mol$_{\rm f.u.}$) and the band Pauli paramagnetic
susceptibility ($\times10^6$).}
\begin{ruledtabular}
\begin{tabular}{ccccc}
& NaOs$_2$O$_6$ & KOs$_2$O$_6$ & RbOs$_2$O$_6$ & CsOs$_2$O$_6$ \\
\hline
Total   & 9.13  &  9.64  & 9.96 & 10.58 \\
\hline
$A$  & 0.019 & 0.020 & 0.030 & 0.048 \\
Os & 4.542 &  4.783 & 4.927 & 5.191  \\
O & 2.578 &   2.751 & 2.864 & 3.070 \\
\hline
$\gamma_{\bf b}$ & 10.76 & 11.36 & 11.74 & 12.46 \\
$\chi_{\bf b}$ & 1.81 & 1.90 & 1.95 & 2.04 \\
\end{tabular}
\end{ruledtabular}
\end{table}

\begin{table}
\caption{\label{tab3} $N(E_{\rm F})$ as a function of pressure, for
KOs$_2$O$_6$ (in states/eV unit cell).}
\begin{ruledtabular}
\begin{tabular}{ccc}
$\Delta V/V$ & P (GPa) & $N(E_{\rm F})$ \\
0\% & 0 & 9.64 \\
0.5\% & 0.583 & 9.47 \\
1\% & 1.166 & 9.33 \\
3\% & 3.498 & 7.24 \\
\end{tabular}
\end{ruledtabular}
\end{table}

\begin{figure}
\begin{center}
\includegraphics[width=\hsize]{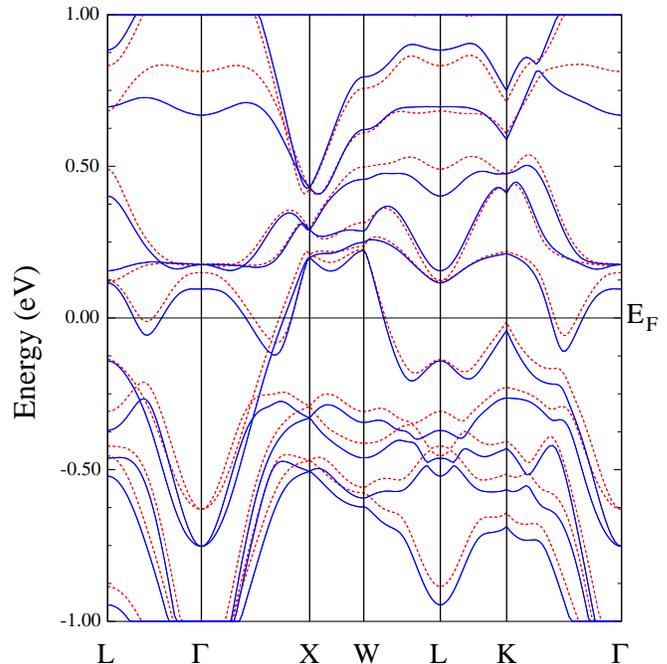}
\end{center}
\caption{\label{fig3} (Color online)
Band structure of RbOs$_2$O$_6$ for energies
within 1 eV to $E_{\rm F}$.
Red-dotted curves: no pressure; solid-blue curves: hydrostatic
pressure such that
$\Delta a/a=-2$\%. Clearly, the the van Hove
singularity is pushed away from $E_{\rm F}$.}
\end{figure}

The effect of hydrostatic pressure is basically
to push the eigenvalues around $E_{\rm F}$ downward. We
illustrate this in the case
of RbOs$_2$O$_6$ in Fig.~\ref{fig3}, where we show a close look at
the bands around $E_{\rm F}$ both for a sample under zero
pressure
and a sample under simulated pressure such that the change in the
lattice constant is $\Delta a/a=-2$\%.
While this corresponds to a relatively large pressure,
it shows clearly the effect on the vHS, pushing it away
from $E_{\rm F}$. This naturally
leads to a decrease of $N(E_{\rm F})$. As a further,
quantitative illustration,
in Table~\ref{tab3}
we list $N(E_{\rm F})$ for
various pressures in the case of KOs$_2$O$_6$.

\begin{figure}
\begin{center}
\includegraphics[width=.75\hsize]{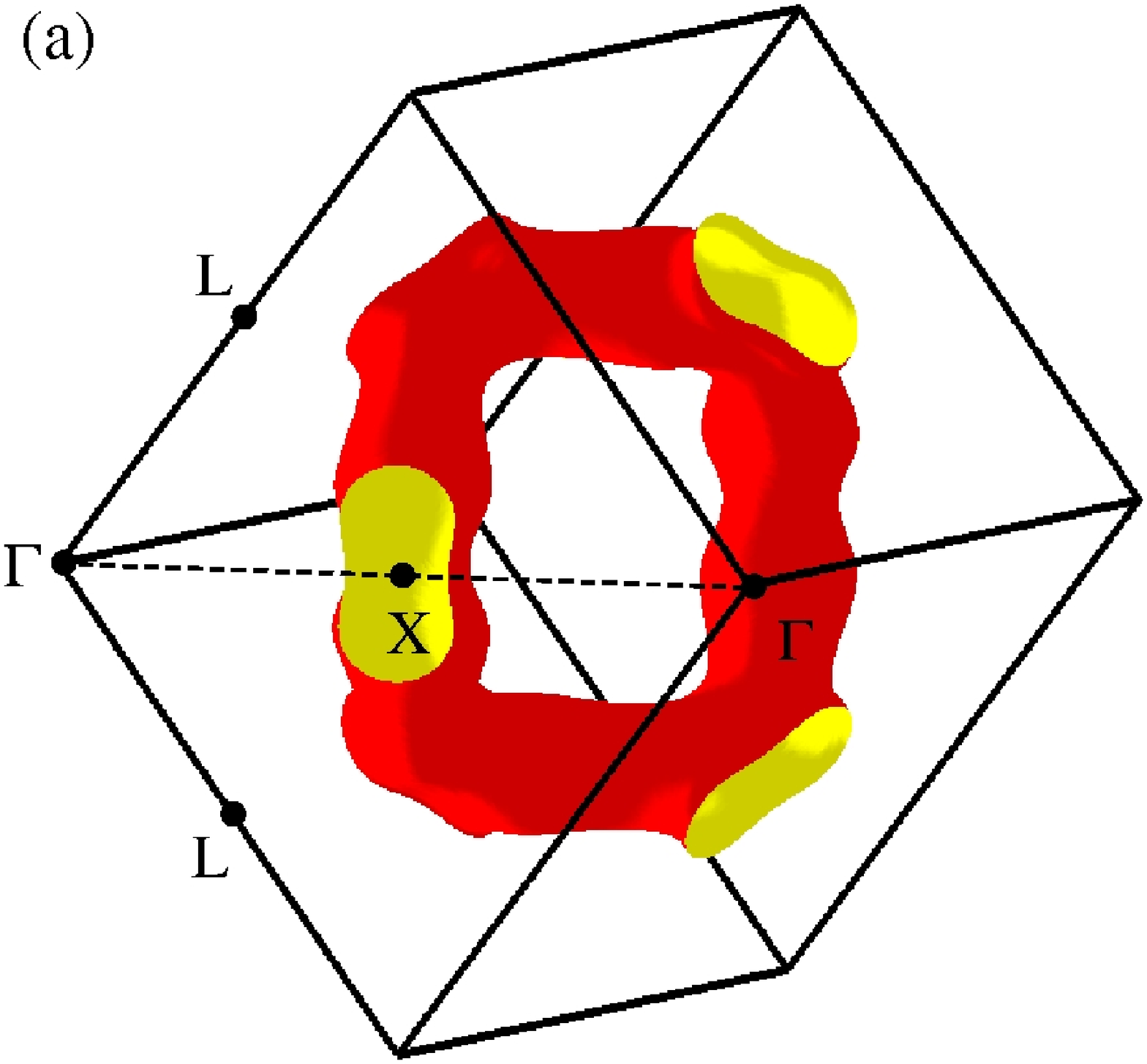}
\end{center}
\begin{center}
\includegraphics[width=.75\hsize]{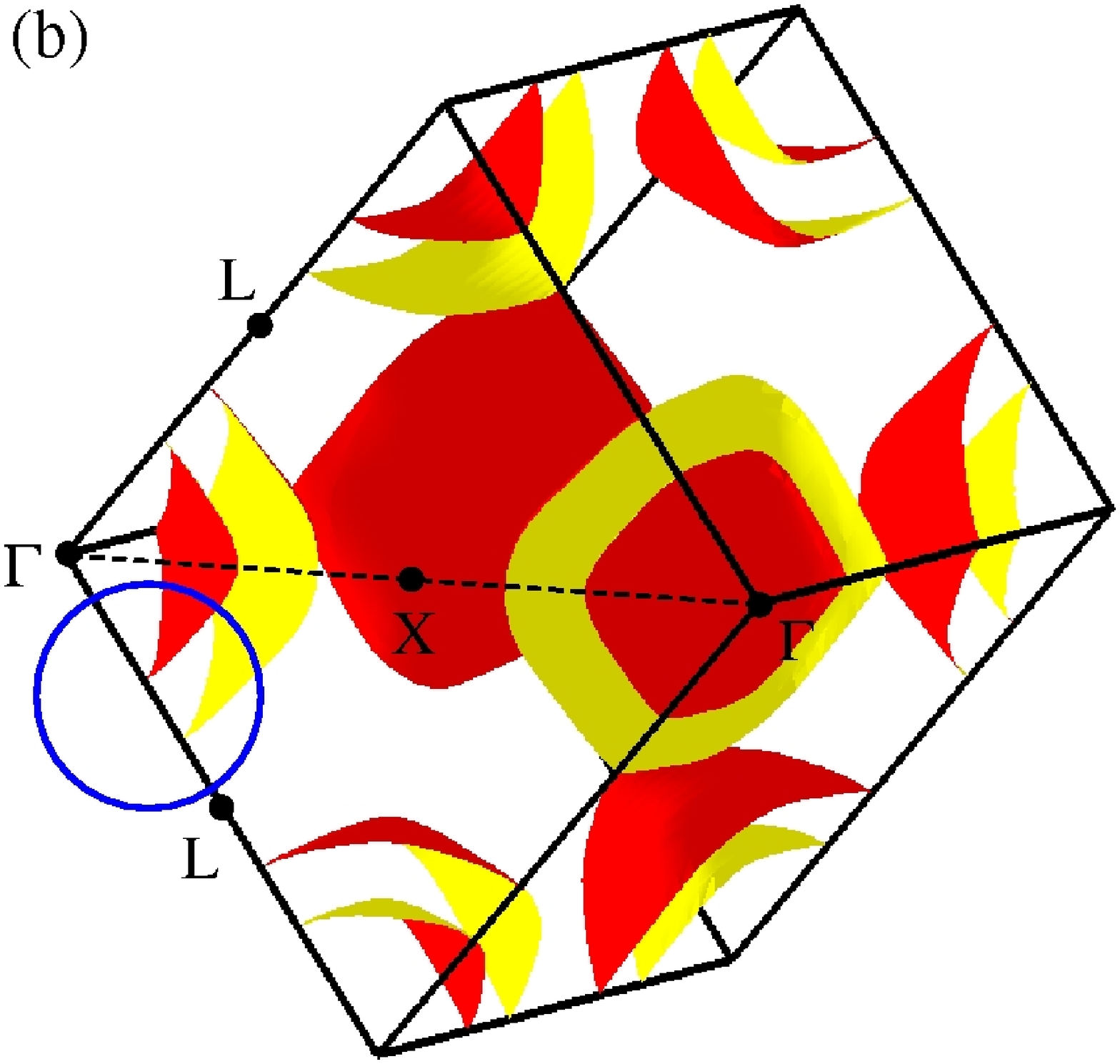}
\end{center}
\begin{center}
\includegraphics[width=.75\hsize]{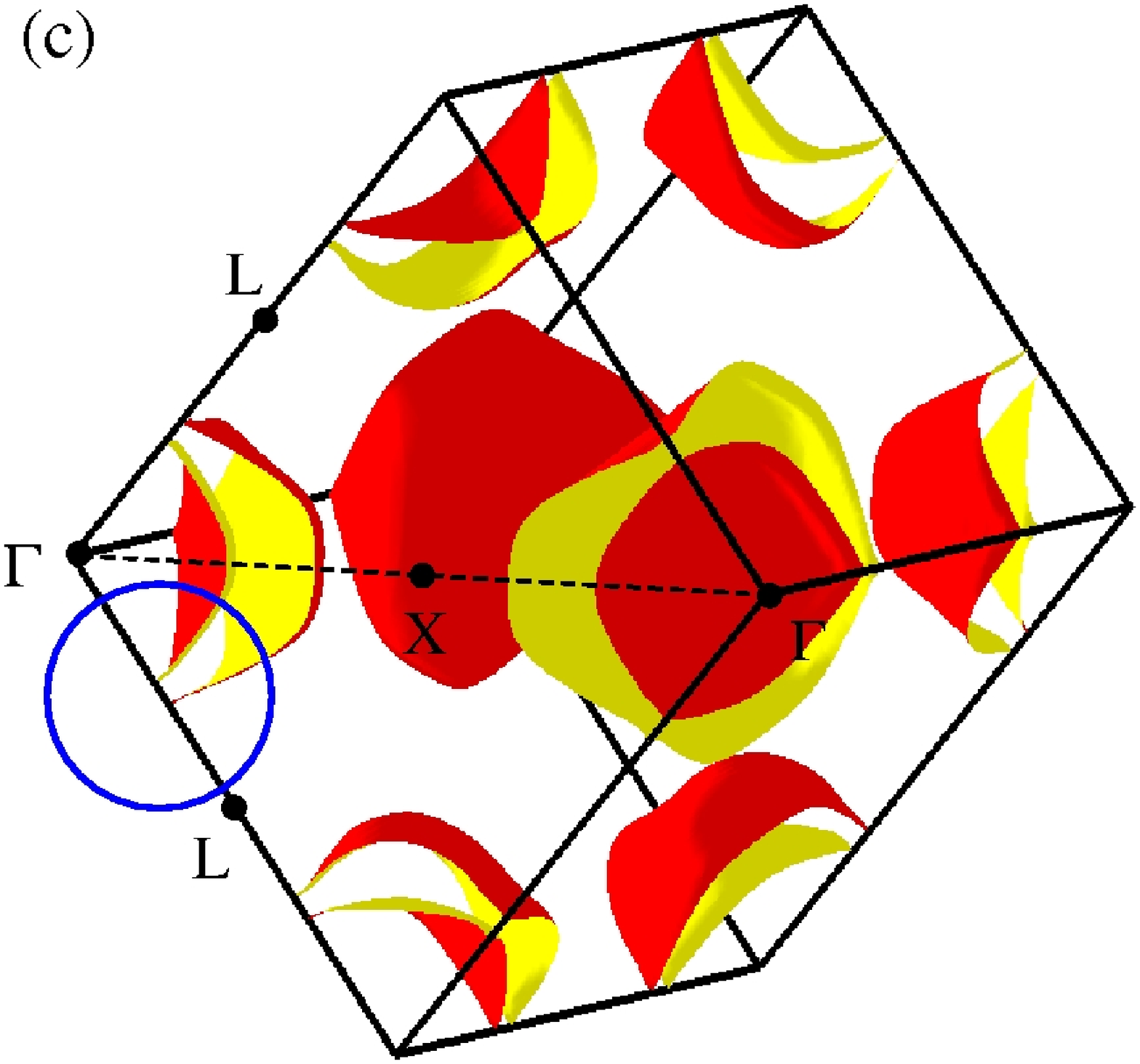}
\end{center}
\caption{\label{fig4} (Color online)
Fermi surface sheets of the $\beta$-pyrochlore
Os oxides, plotted in the reciprocal unit cell. The ``outer'' surfaces
of the sheets are colored red, while the ``inner'' ones are
yellow. (a) The hole-like
tubular network, which
does not change noticeably under change
of the alkali metal. (b) The closed shell surfaces around the
$\Gamma$ point for NaOs$_2$O$_6$. No constriction along the
$\Gamma$-$L$ line is seen (cf. circle). (c)
The constriction near the center of the $\Gamma$-$L$ line,
indicated by the circle, in the
case of CsOs$_2$O$_6$ is clear.
}
\end{figure}

As one may expect,
the above effects are reflected in the topology of the Fermi
surface. This is of general interest because
of its direct relation to electronic properties and
the possible effect of the vHS on quasi-particle lifetimes.
The Fermi surface consists of two closed electron-like
sheets centered at the
$\Gamma$ point, and a third hole-like sheet giving rise to a
tubular network. These Fermi surface
sheets are shown in Fig.~\ref{fig4}, plotted for clarity
in the reciprocal unit cell. The tubular network
actually does not present any major difference
among the compounds considered; as a typical example we show the
case of CsOs$_2$O$_6$ in Fig.~\ref{fig4}(a).
In contrast, the closed shells exhibit a clear difference near the
midpoint of the $\Gamma$-$L$ line, where the vHS is located.
Indeed, in the case of
NaOs$_2$O$_6$, in Fig.~\ref{fig4}(b),
the two shells show no
narrowing of the distance at this point, while the narrowing
is obvious in the case of CsOs$_2$O$_6$, as shown
in Fig.~\ref{fig4}(c). As discussed above, this is due to the
closeness of the vHS to $E_{\rm F}$ in the latter case.
The topology of the Fermi surface is also relevant to the
superconducting gap.
In relation to this,
we note that the multi-band character of the Fermi surface
and the different
symmetry of its sheets may be of significance to
some of the experimental results on the
$\beta$-pyrochlores. As pointed out above,
Koda and co-workers interpret their results on the
linear field dependence of the penetration depth,
$\lambda$, as suggesting a
non-conventional pairing mechanism in KOs$_2$O$_6$,\cite{koda04}
possibly mediated by magnetic fluctuations.
However, in MgB$_2$, which is a phonon-mediated superconductor,
$\lambda$ also exhibits such a linear dependence on the applied
field.\cite{ohishi03} In the case of MgB$_2$ this arises
because of its two-gap nature, which in turn is due to the
particular character of its Fermi surface sheets, which
we find akin to the present case to some extent.

\subsection{Superconducting parameters}

In the following, we examine the ability of a phonon-mediated
pairing scenario to account for the experimental evidence,
and, more particularly,
the effects of alkali metal substitution
and of pressure. To this end,
we estimate the electron-phonon coupling
constant $\lambda_{\rm ep}$ within the McMillan-Hopfield
framework,\cite{mcmillan68,hopfield69} and the crude rigid-muffin-tin
approximation.\cite{gaspari72,pettifor77} The spherically averaged
Hopfield parameter can be written\cite{skriver85}
\begin{equation}
\eta_i=\sum_lM^2_{i_{l,l+1}}{{2l+2}\over{(2l+1)(2l+3)}}\left[
{{N_{i_l}(E_{\rm F})N_{i_{l+1}}(E_{\rm F})}\over{N(E_{\rm F})}}
\right],
\end{equation}
where $N_{i_l}(E_{\rm F})$ is the $l$-angular momentum DOS
projected on the muffin-tin sphere of atom $i$;
$M_{i_{l,l+1}}=-\phi_{i_l}\phi_{i_{l+1}}[(D_{i_l}-l)
(D_{i_{l+1}}+l+2)+
(E_{\rm F}-V_i)R_i^2]$ is an electron-phonon matrix element, in terms
of the logarithmic derivatives
($D_{i_l}$) and the partial wave amplitudes
($\phi_{i_l}$), both evaluated at
$E_{\rm F}$ and at the muffin-tin
radius ($R_i$); $V_i$ is the one-electron potential at $R_i$.
Then we have $\lambda_{\rm ep}=\sum_i\eta_i/\bar M\langle
\omega_i^2\rangle$, where $\bar M$ is the average
mass.\cite{comment2}
As is common, the average phonon frequency
is estimated in terms of the Debye temperature as
$\langle\omega^2\rangle^{1/2}=0.69\,\Theta_{\rm D}$.

Regarding the Debye temperatures, there is
unfortunately no experimental information for
KOs$_2$O$_6$ and CsOs$_2$O$_6$. In the case of
RbOs$_2$O$_6$, recent measurements
suggest that the specific heat does not follow the usual
$\sim T^3$ Debye model,\cite{comment3}
so that it is not clear what effective $\Theta_{\rm D}$ is
appropriate in our case. With this caveat,
in our estimates we use the value
$\Theta_{\rm D}=285$ K at $T=6$ K obtained if a 
parametrization of the
heat capacity in terms of a $T$ dependent $\Theta_{\rm D}$
is enforced.\cite{comment3}
In Table~\ref{tab4}, we present our results for
the $\eta_i$ and $\lambda_{\rm ep}$ for the different
compounds.\cite{comment4}

\begin{table}
\caption{\label{tab4} Electron-phonon coupling parameters of the
Os oxide $\beta$-pyrochlores in the RMTA approximation,
with $\Theta_{\rm D}=285$ K.
Hopfield parameters in eV/\AA$^2$.}
\begin{ruledtabular}
\begin{tabular}{ccccc}
Compound & $\eta_{\rm A}$
&$\eta_{\rm Os}$  & $\eta_{\rm O}$  &
$\lambda_{\rm ep}$  \\
\hline
NaOs$_2$O$_6$ & $3\times 10^{-6}$
 & 0.421 & 0.126 & 0.84 \\
KOs$_2$O$_6$ & $1\times10^{-4}$
 & 0.439 & 0.133 & 0.85 \\
RbOs$_2$O$_6$ & $2\times10^{-4}$
 & 0.449 & 0.137 & 0.80 \\
CsOs$_2$O$_6$ & $5\times10^{-4}$
 & 0.464 & 0.145 & 0.77 \\
\end{tabular}
\end{ruledtabular}
\end{table}
                                                                            
\begin{table*}
\caption{\label{tab5} Superconducting parameters of the
Os oxide $\beta$-pyrochlores. $\mu^*$ is the Coulomb
pseudopotential; $I$ and $S$ are the Stoner parameter and
susceptibility enhancement factor, respectively,
and $\mu_{\rm sp}$
is the electron-spin coupling constant; $T_c'$ denotes the
calculated critical temperature with $\mu_{\rm sp}=0$. }
\begin{ruledtabular}
\begin{tabular}{cccccccc}
Compound & $\mu^*$ & $I$ (eV) & $S$ & $\mu_{\rm sp}$ &
$T_c'$ (K) & $T_c^{\rm exp}$ (K) & $T_c$ (K) \\
\hline
NaOs$_2$O$_6$ & 0.088 & 0.1166 & 2.14 & 0.054 & 10.9 & - & 7.0 \\
KOs$_2$O$_6$ &  0.091 & 0.1166 & 2.28 & 0.064 & 11.0 & 9.6 & 6.4  \\
RbOs$_2$O$_6$ & 0.093 & 0.1162 & 2.37 & 0.070 & 9.7 & 6.3 & 5.0 \\
CsOs$_2$O$_6$ & 0.096 & 0.1155 & 2.57 & 0.084 & 8.8 & 3.3 & 3.6  \\
\end{tabular}
\end{ruledtabular}
\end{table*}

The critical temperature is subsequently calculated with the
McMillan-Allen-Dynes equation\cite{bennemann72,allen75}
\begin{equation}
T_c={{\langle\omega^2\rangle^{1/2}}\over{1.2}}
\exp\left[-{{1.04(1+\lambda_{\rm ep}+\mu_{\rm sp})}\over
{\lambda_{\rm ep}-(\mu^*+\mu_{\rm sp})(1+
0.62\lambda_{\rm ep})}}\right],
\end{equation}
where $\mu^*$ is the Coulomb pseudopotential and $\mu_{\rm sp}$
is an effective electron-spin coupling constant. Note that,
lacking the knowledge of the Eliashberg function
$\alpha^2F(\omega)$, in lieu of the logarithmic average
$\omega_{\ln}$ we take the average phonon frequency, which we estimate
in terms of $\Theta_{\rm D}$, as indicated above.\cite{comment5}
The Coulomb pseudopotential can be estimated through
$\mu^*=0.26n(E_{\rm F})/[1+n(E_{\rm F})]$, where
$n(E_{\rm F})$ is the DOS at $E_{\rm F}$ per eV and
atom.\cite{bennemann72} To estimate $\mu_{\rm sp}$ we use
the expression derived by Doniach and
Engelsberg
$\mu_{\rm sp}\simeq 3IN(E_{\rm F})\ln\{1+0.03 IN(E_{\rm F})/
[1-IN(E_{\rm F})]\}.$\cite{doniach66,comment6}
The Stoner parameter in this
expression, $I$, is calculated following the band formulation
of Gunnarsson\cite{gunnarsson76} and Brooks
{\it et al.}\cite{brooks87}
within spin-density-functional theory. More specifically, we
have $I=\sum_in_iI_i,$ where $n_i$ is the number of atoms of
type $i$, and $I_i$ the atomic Stoner parameter
written as $I_i=\sum\hat n_{i_{ll'}}J_{i_{ll'}}$. Here
$\hat n_{i_{ll'}}=N_{i_l}(E_{\rm F})N_{i_{l'}}(E_{\rm F})/
N_i^2(E_{\rm F})$
and $J_{i_{ll'}}=\int dr |K(r)| \phi_{i_l}^2(r)
\phi^2_{i_{l'}}(r)$ with,
again, the partial wave amplitudes $\phi_{i_l}$ calculated at
$E_{\rm F}$. The exchange-correlation kernel $K$ used is
the one given by Gunnarsson.\cite{gunnarsson76} In our case,
the alkali atom contribution is completely negligible and only
the diagonal $l=1$ term in O and the diagonal
$l=2$ in Os contribute
because of the dominance of the respective partial
DOS at $E_{\rm F}$ (see, e.g., Ref.~\onlinecite{brooks87}).
We note that our calculations
are done taking spin-orbit coupling into account.

Our results for the superconducting
parameters, as well as for $I$ and the Stoner enhancement factor
$S=1/[1-IN(E_{\rm F})/2]$, are given in
Table~\ref{tab5}. We note first
that our calculated $T_c$ (last column) follows well
the experimental trend, although the range of the experimental
$T_c$'s is noticeably larger. Clearly, however, a more
refined calculation of $\lambda_{\rm ep}$ 
based on, e.g., the
frequency dependence of the Eliashberg function $\alpha^2F$,
can easily account for the difference
of a fraction to a few K between our results and
experiment seen in Table~\ref{tab5}.
In this regard, Kune\v s and co-workers\cite{kunes04} have
found that the alkali ions in these materials possess a
varying degree of anharmonicity in the potential,
which would add to
the difference in their $T_c$'s.
Secondly, we note the significant role of
spin fluctuations. Indeed, with $\mu_{\rm sp}=0$ the
predicted $T_c$ (given as $T_c'$) is $\sim$72\% (KOs$_2$O$_6$)
to $\sim$144\% (CsOs$_2$O$_6$) higher. Thus,
although the calculated Stoner enhancement factors
($2<S<3$) indicate that these systems
are not close to a ferromagnetic instability, they are sufficient
to produce a significant electron-spin coupling.

Finally, we have calculated the superconducting parameters
of KOs$_2$O$_6$ (again with $\Theta_{\rm D}=285$ K)
under the simulated effect of pressure, to study the initial change
of $T_c$ with pressure. We have considered pressures up to
1.166 GPa, which is of the order of the pressures used in the
experimental report by Muramatsu {\it et al.}\cite{muramatsu04}
Our results are given in Table~\ref{tab6}. As in
experiment,\cite{khasanov04,muramatsu04}
we see that $T_c$ increases with pressure, although
$\lambda_{\rm ep}$ decreases.
The main reason is that both
$\mu^*$ and $\mu_{\rm sp}$ also decrease, and more
importantly the latter than the former. Hence, the initial
increase of
$T_c$ with pressure appears to be due mainly to a decrease
of spin fluctuations, driven by the decrease of $N(E_{\rm F})$
(the Stoner parameter $I$ is almost unchanged).
Comparing our results with the ratio
$T_c/T^0_c=1.04$ found at 0.56 GPa by Muramatsu and
collaborators,\cite{muramatsu04} we see that the change in
$T_c$ is of the same order and is
essentially accounted for. Indeed, if the main cause
were to be phononic, i.e., a change in $\Theta_{\rm D}$,
the latter would have to
decrease with pressure, contrary to any
likelihood.\cite{comment7}
Again, we surmise that a more
refined calculation can readily account for the somewhat
steeper initial increase of $T_c$ observed.
In their experimental study of the change of $T_c$ with
pressure in the case of RbOs$_2$O$_6$,
Khasanov and
co-workers had already concluded that the
observed positive slope at low pressures
must arise mainly from electronic
contributions, as opposed to phononic ones, although the mechanism
behind the effect was still an open
question.\cite{khasanov04} 

We note that the very recent report by Muramatsu and
collaborators\cite{muramatsu05} shows that
in the present family of superconductors, after reaching a
maximum value, with increasing pressure
$T_c$ will gradually decrease, falling below its ambient pressure
value, until superconductivity is eventually suppressed.
To make predictions of $T_c$ at those high pressures,
however, it would be necessary to have a minimum information on 
the phonon spectra and how they are affected by pressure.
We do not know what would be
a sensible value of $\Theta_{\rm D}$ to make such estimates of $T_c$
within our approach. However, we can try to understand what
happens as follows. If the Gr\"uneisen parameter for
KOs$_2$O$_6$ is around 1.8, which is a rough average for
most substances,\cite{gschneider64} then for a pressure $P=3.5$ GPa
(corresponding to $-\Delta V/V\simeq3$\%),
$\Theta_{\rm D}\simeq 301$ K.
The calculated superconducting parameters ($\mu^*=0.085$,
$\mu_{\rm sp}=0.048$, and $\lambda_{\rm ep}=0.738$) then
would lead to $T_c/T_c^0\simeq 0.88$ (against $\simeq 0.66$ in
experiment). This result suggests that the decrease of $T_c$ at
higher pressures could be understood as reflecting
the fact that in that regime 
the phononic properties become dominant.

It is remarkable, given the approximate nature of our
estimate of $\lambda_{\rm ep}$ and $T_c$, that our results
account rather well for the behavior of $T_c$ under substitution
of the alkali metal and under applied pressure. We believe this
brings strong support for the phonon-mediated pairing
scenario.
To understand more fully the properties of the $\beta$-pyrochlore
Os oxides, however, further investigation will be required,
experimentally and theoretically.
For instance, the different nuclear spin-lattice relaxation
rates of the alkali ions in KOs$_2$O$_6$ and RbOs$_2$O$_6$
remain to be clarified. This could be partly due to the
rapid variation of the DOS close to $E_{\rm F}$. Indeed,
Fig.~\ref{fig2} shows that $N(E_{\rm F})$ can change by as much
as 50\% within an energy range of $\pm$25 meV (more so the
heavier the alkali ion).
A further contribution may come from the
changing character of the alkali atom DOS at $E_{\rm F}$.
We find that
its $s$ character falls from 73\% in NaOs$_2$O$_6$,
to 11\% in
CsOs$_2$O$_6$, passing by 31\% in KOs$_2$O$_6$ and
23\% in RbOs$_2$O$_6$ (at the same time its $p$ character rises
from 16\% in NaOs$_2$O$_6$,
to 55\% in
CsOs$_2$O$_6$, passing by 31\% in KOs$_2$O$_6$ and
38\% in RbOs$_2$O$_6$).

Furthermore, the origin of the unusual behavior of
the resistivity in KOs$_2$O$_6$ at low $T$ is also
not understood,
nor is the rather large specific heat mass enhancements
$\gamma_{\rm exp}/\gamma_{\rm b}$ in all these materials.
The electron-phonon and electron-spin coupling constants
obtained above are clearly insufficient to account for the
observed enhancements of 3--4. It is possible that the
vHS near $E_{\rm F}$ and the nesting exhibited by the Fermi
surface\cite{comment8} contribute to both the resistivity
and the specific heat. The observed
enhancements suggest, in our view, that the unusual
non-Debye behavior at low temperature of the specific heat
mentioned above
is a generic property.

\begin{table}
\caption{\label{tab6} Superconducting parameters as a function of
pressure for KOs$_2$O$_6$ ($\Theta_{\rm D}=285$ K).
$T_c^0$ corresponds to 0 pressure. }
\begin{ruledtabular}
\begin{tabular}{cccccc}
Pressure (GPa) & $\mu^*$ & $S$ & $\mu_{\rm sp}$ & $\lambda_{\rm ep}$
& $T_c/T^0_c$ \\
\hline
0 & 0.091 & 2.28 & 0.064 & 0.852 & 1 \\
0.583 & 0.090 & 2.23 & 0.060 & 0.845 & 1.03 \\
1.166 & 0.089 & 2.19 & 0.057 & 0.841 & 1.05 \\
\end{tabular}
\end{ruledtabular}
\end{table}

\begin{acknowledgments}

We thank Lin-Hui Ye, S. H. Rhim, J. B. Ketterson, and
W. Halperin for helpful discussions. We are also grateful to
B. Barbiellini and G. Grimvall for their suggestions and to
B. Battlogg and M. Br\"uhwiler for sharing data with
us prior to publication.
This work was supported by the Department of
Energy (under grant No. DE-FG02-88ER 45372/A021
and a computer time grant at the
National Energy Research Scientific Computing Center).

\end{acknowledgments}


\begin{thebibliography}{99}
\bibitem{comment1} This result calls for similar measurements on
the other Os oxide $\beta$-pyrochlores, not least because,
in addition, a second peak is observed in the $C/T$ curve, below
the superconducting $T_c$, which could indicate a further first-order
or $\lambda$-type phase transition.
\bibitem{comment6} As Doniach and Engelsberg
(Ref.~\onlinecite{doniach66}), we use a momentum cutoff
$p/p_{\rm F}=0.6$.
\bibitem{comment7} As noted above (cf. Ref.~\onlinecite{comment4}),
$\Theta_{\rm D}^{\rm e}$ increases with decreasing lattice
constant (positive chemical pressure),
thus indicating a positive Gr\"uneisen constant.
\bibitem{comment8} In Ref.~\onlinecite{kunes04}, a calculation of
the bare band susceptibility along the $\Gamma$-$K$ direction
is reported for KOs$_2$O$_6$. Although it appears not to suggest
electronic instabilities, it does show clear nesting effects.
We note that the constriction between the two closed Fermi surfaces
discussed above
may have effects on the nesting; in particular, it
may degrade any nesting along the [111] direction in the case of
the heavier alkali metals.
\end{thebibliography}
\end{document}